\documentstyle[12pt]{article}

\begin{document}
\topmargin 0pt
\oddsidemargin 7mm
\headheight 0pt
\topskip 0mm

\addtolength{\baselineskip}{0.40\baselineskip}

\hfill SNUTP 94-110

\hfill  hep-th/9502114

\begin{center}
\vspace{0.5cm}
{\large \bf Stability Analysis of the Dilatonic Black Hole in Two Dimensions}
\end{center}

\vspace{1cm}

\begin{center}
Won Tae Kim$^{(a)}$ and Julian Lee$^{(b)}$ \\
{\it Center for Theoretical Physics and Department of Physics, \\
Seoul National University, Seoul 151-742, Korea}

\vspace{0.5cm}

Young-Jai Park$^{(c)}$

{\it Department of Physics and Basic Science Research Institute, \\
Sogang University, C.P.O. Box 1142, Seoul 100-611, Korea}

\end{center}

\vspace{1cm}

\begin{center}
{\bf ABSTRACT}
\end{center}

We explicitly show that the net number of degrees of freedom in the
two-dimensional dilaton gravity is zero through the Hamiltonian
constraint analysis. This implies that the local space-time dependent
physical excitations do not exist. From the linear perturbation
around the black hole background, we explicitly prove that
the exponentially growing mode with time
is in fact eliminated
outside the horizon.
Therefore, the two-dimensional dilation gravity is essentially stable.

\vspace{2cm}
\hrule
\vspace{0.5cm}
\hspace{-0.6cm}$^{(a)}$ E-mail address : wtkim@phyb.snu.ac.kr \\
\hspace{-0.6cm}$^{(b)}$ E-mail address : lee@phyb.snu.ac.kr \\
\hspace{-0.6cm}$^{(c)}$ E-mail address : yjpark@ccs.sogang.ac.kr

\newpage

\pagestyle{plain}

A black hole solution to the two-dimensional critical string theory
has attracted much interest [1]. It is given by the solution of the
modular invariant $SL(2,R)/U(1)$ gauged Wess-Zumino-Witten(WZW) coset
model of a conformal field theory. The black hole solution can be also
derived by solving the two-dimensional beta-function equations of the string
theory, which is effectively described by the two-dimensional dilaton
gravity at the Lagrangian level [2].
This theory can be used as a toy model in two
dimensions which can be the basic starting point in resolving the interesting
puzzles of gravitational system such as the end point of Hawking radiation and
information loss problem inside the black hole [3,4]. Then it becomes natural
task to investigate the classical stability of the black hole solution for this
model. To determine whether the black hole is stable or not,
the small (linear) perturbation of the classical equations of motion in the
black hole background
was considered in the regular region, which  usually sees the effective
potential [5]. As is well known, if there exists an exponentially
growing mode with time, the black hole is unstable [6].
At first sight, one can understand that there should be no physical
instability, because one can easily see that the degrees of freedom in the
two-dimensional dilaton gravity is zero in the absence of matter fields.
However it was claimed that
the growing mode with time in the gravity sector exists [7].
Therefore,
their claim amounts to the statement that the black hole is unstable
in the absence of matter fields.  More recently, it was conjectured
that the exponentially growing mode with time induced by potential well
should be a gauge artifact simply by counting degrees of freedom [8]
for the gravitational field which is given by ${\rm d(d-3)/2}$
in d dimensions [9].

In this paper, we reconsider the two-dimensional dilaton gravity to
clarify these points and elaborate whether the black hole is really
stable or not.
In particular, we will show that the
two-dimensional static black hole solution is stable against
linear perturbation since the unstable mode related to the global
symmetry of the theory can be eliminated by an appropriate
global coordinate
transformation.
As a guidance, we first enumerate the net number of degrees of freedom
in terms of Hamiltonian constraint analysis where it turns out to be a
null degree. It implies that no local propagating modes exist.
{}From this point of view, we try to find what is the possible perturbation of
black hole solution, and show that the only possible perturbations correspond
to the variations of
the three global parameters, one is the black hole mass and the others are
the position coordinates of black hole.
Next we shall apply the conventional linear
perturbation method around the black hole background to explicitly see the
behavior of the classical perturbation modes.
We then identify the
growing mode with time with the variations of
the global parameters by comparing the above results.
We conclude that the exponentially growing mode with time
is eliminated by choosing a suitable position coordinate, which means
that the two-dimensional dilatonic black hole is stable against the
linear perturbation.

Let us now start with the two-dimensional dilaton gravity defined by [3]
\begin{equation}
\label{eq:lag}
    S_D = \int d^2 x \sqrt{G} e^{-2\phi}
                 [ R + 4 g^{\mu\nu} \nabla_{\mu} \phi \nabla_{\nu} \phi
                   + 4 \lambda^2],
\end{equation}
where $G=-{\rm det} g_{\mu \nu}$ and $\phi$ is a dilaton field,
and $\lambda$ is a cosmological constant.

Following the Arnowitt-Deser-Misner(ADM) formulation, the
two-dimensional $g_{\mu \nu}$ is written by
\begin{eqnarray}
\label{eq:m}
    g_{\mu \nu} = \gamma
                  \left( \begin{array}{cc}
                          -N_0^2+N_1^2 & N_1 \\
                          N_1          & 1
                         \end{array}
                  \right),
\end{eqnarray}
where $N_0$ and $N_1$ are lapse and shift functions respectively, and we
factor out the conformal factor $\gamma$ [10].

Then the action (\ref{eq:lag}) can be rewritten by the first order form as
follows
\begin{equation}
\label{eq:flag}
S_D = \int d^2 x (\pi_{\phi} \partial_0 \phi
                + \pi_{\gamma} \partial_0 \gamma - N_0 G_0 - N_1 G_1),
\end{equation}
where the generators of reparametrization are
\begin{eqnarray}
\label{eq:c}
G_0 &=& 2 \gamma e^{2\phi} (\pi_\phi + 2 \gamma\pi_\gamma)\pi_\gamma
           - e^{-2\phi} ((\partial_1 \phi)^2 + \lambda^2), \nonumber \\
G_1 &=& - \pi_\gamma \partial_1 \gamma + \pi_\phi \partial_1 \phi,
\end{eqnarray}
and $\pi_{\phi}$, $\pi_{\gamma}$ are canonical momenta with respect to
the dilaton and conformal factor $\gamma$ respectively.
Two primary constraints for auxiliary fields are $\pi_{N_0}$ and
$\pi_{N_1}$, which are fully first class constraints [11].
Two secondary constraints $G_0$, $G_1$ corresponding to the Virasoro
constraints satisfy the closed algebra under Poisson brackets,
\begin{eqnarray}
\label{eq:ca}
\{ G_0(x), G_0(y) \} &=& ( G_1(x) + G_1(y) ) \partial_1 (x^1-y^1), \nonumber\\
\{ G_1(x), G_1(y) \} &=& ( G_1(x) + G_1(y) ) \partial_1 (x^1-y^1), \\
\{ G_0(x), G_1(y) \} &=& ( G_0(x) + G_0(y) ) \partial_1 (x^1-y^1). \nonumber
\end{eqnarray}
Therefore, the true physical degree of freedom is zero because
the graviton and dilaton degrees of freedom are four,
while there are four first-class constraints,
{\it{i.e.}}, $(3+1)-(2+2)=0$.
As a natural result, the local excitations are absent.
However, there are still global degrees of freedom,
which will be considered later.

The equations of motion for the graviton and dilaton are given by
\begin{eqnarray}
\label{eq:eom}
R_{\mu\nu} &+& 2 \nabla_\mu \nabla_\nu \phi = 0, \nonumber \\
\Box \phi &-& 2 (\nabla\phi)^2 + 2 \lambda^2 =0.
\end{eqnarray}
The classical theory is most easily analyzed in the conformal gauge
as follows
\begin{equation}
\label{eq:cg}
ds^2 = 2 g_{+-} dx^+ dx^-, ~~~~~~g_{+-} = -\frac{1}{2} e^{2\rho(x^+, x^-)},
\end{equation}
where $x^{\pm}=x^0 \pm x^1$, and the non-vanishing Christoffel symbols are
$\Gamma^+_{++}=2\partial_+ \rho$ and $\Gamma^-_{--}=2\partial_- \rho$.
In the conformal gauge, we have the equation of motion
\begin{equation}
\label{eq:kru}
\partial_-\partial_+(\rho - \phi) = 0,
\end{equation}
which yields
\begin{equation}
\label{eq:tran}
\rho(x) = \phi(x)+\frac{1}{2} (w_+(x^+)+w_-(x^-))
\end{equation}
with arbitrary functions $w_\pm(x^\pm)$. However, in conformal gauge there is
still a residual gauge transformation of the form
\begin{equation}
\label{eq:ww}
\rho'(x')=\rho(x) - \frac{1}{2}\left(
\ln (\frac{\partial {x'}^+}{\partial x^+})
+ \ln (\frac{\partial {x'}^-}{\partial x^-})
                          \right).
\end{equation}
Therefore we can fix the residual gauge symmetry such that the functions
 $w_\pm$ vanishes. Our calculations will be
done in this Kruskal gauge. This gauge is quite a rigid one and only possible
coordinate transformation, which preserves the condition (\ref{eq:kru}),
is the
 global transformation $x^\pm  \to x^\pm + a^\pm$ with some constant
 $a^\pm$, which
is just
a translation. We will see later that the exponentially growing modes are
directly related to it.
In this gauge, the black hole solution is given by [3]
\begin{equation}
\label{eq:bhs}
e^{-2 \bar{\rho}(x^+, x^-)} = e^{-2\bar{\phi}(x^+,x^- )}
                            = \frac{m}{\lambda} - \lambda^2
                              (x^+ -x_0^+)(x^- -x_0^-),
\end{equation}
where $m$ is an integration constant which turns out to be a black
hole mass [1].

Note that the event horizon of the black hole solution (\ref{eq:bhs}) is
$x^+_H =x_0^+$ or $ x^-_H=x_0^-$. The arbitrariness is due to the
above mentioned translational
symmetry, and we may freely set the black hole position
$x_0^\pm=0$ by a global transformation $x^\pm \to x^\pm + x_0^\pm$ in eq. (11).
This black hole solution (\ref{eq:bhs}) is exact without any approximations
because differential equations (\ref{eq:eom}) are exactly solvable.
Furthermore, the system has no local degrees of freedom, and
the black hole solution
describes the global geometry. In this two-dimensional dilatonic gravity,
the only possible perturbation around the static background is due to the
variation of three global parameters, which are black hole mass $m$
and the black hole position $x_0^{\pm}$.
To make it explicit, we can vary these parameters in (\ref{eq:bhs}) to get
\begin{eqnarray}
\label{eq:delphipm}
\phi(x^+ \hspace{-0.3cm}&,&\hspace{-0.3cm}x^-  ; m+\delta m, x_0^{\pm}
                   + \delta x_0^{\pm})
                     - \bar{\phi} (x^+,x^-;m, x_0^{\pm}) \nonumber \\
   & =& \delta m \frac{\partial}{\partial m} \bar{\phi} (x^+,x^-;m,
x_0^{\pm})\nonumber \\
   &  +& \delta x_0^+ \frac{\partial}{\partial x_0^+}
         \bar{\phi} (x^+,x^-;m, x_0^{\pm})
     + \delta x_0^- \frac{\partial}{\partial x_0^-}
         \bar{\phi} (x^+,x^-;m, x_0^{\pm})
                            + \cdots  \nonumber \\
    &=& \frac{ -\frac{\delta m}{2\lambda} -\frac{\lambda^2}{2}
             (\delta x_0^-  x^+ +\delta x_0^+ x^-)}
            {\left[\frac{m}{\lambda} -\lambda^2 (x^+ -x_0^+)(x^-  -
x_0^-)\right] }+\cdots ,
\end{eqnarray}
where the dots mean higher order terms.
This relation will be used later to identify the perturbation modes.

Let us now study a small perturbation in terms of the conventional linear
perturbation method around the classical black hole solution [5].
The result should be consistent with eq. (\ref{eq:delphipm}) if we identify
some integration constants, which arise from solving the perturbed differential
equations, with known quantities.
The linear perturbed fields around the static black
hole solution $(\bar{g}_{\mu\nu},\bar{\phi})$ are defined by
\begin{eqnarray}
\label{eq:lp}
g_{\mu\nu}(x^+,x^-) &=& \bar{g}_{\mu\nu}(x^+,x^-) + h_{\mu\nu}(x^+,x^-),
                        \nonumber \\
   \phi(x^+,x^-)    &=& \bar{\phi}(x^+,x^-) + \delta\phi(x^+,x^-).
\end{eqnarray}
Then, the linear perturbation leads to the perturbed equations of
motion as follows
\begin{eqnarray}
\label{eq:lpeom}
\delta R_{\mu\nu} \hspace{-0.3cm}&+&\hspace{-0.3cm} 2( \bar{\nabla}_\mu
\bar{\nabla}_\nu \delta \phi
                       - \delta \Gamma_{\mu\nu}^{\alpha}
                       \bar{\nabla}_\alpha \bar{\phi} )  = 0,
                       \nonumber \\
h^{\mu\nu}\hspace{-0.3cm} &(&\hspace{-0.3cm}-\bar{\nabla}_\mu \bar{\nabla}_\nu
\bar{\phi}
            + 2 \bar{\nabla}_\mu \bar{\phi}
              \bar{\nabla}_\nu \bar{\phi} )
   + \bar{g}^{\mu\nu}
       (\bar{\nabla}_\mu \bar{\nabla}_\nu \delta \phi
   - 4 \bar{\nabla}_\mu \bar{\phi} \bar{\nabla}_\nu \delta \phi
   -  \delta \Gamma^{\alpha}_{\mu\nu} \bar{\nabla}_\alpha \bar{\phi}) = 0,
\end{eqnarray}
where the upper bars represent the background quantities, and
$\delta R_{\mu\nu}$, and $\delta\Gamma^{\alpha}_{\mu\nu}$ are given by
\begin{eqnarray}
\delta R_{\mu\nu} &=& \frac{1}{2}
           ( - {\bar{\nabla}}^2 h_{\mu\nu}
            - \bar{\nabla}_{\mu} \bar{\nabla}_{\nu} h^{\alpha}_{\alpha}
            + \bar{\nabla}^{\alpha} \bar{\nabla}_{\mu} h_{\nu\alpha}
            + \bar{\nabla}^{\alpha} \bar{\nabla}_{\nu} h_{\mu\alpha} ),
            \nonumber \\
\delta \Gamma^{\alpha}_{\mu\nu} &=& \frac{1}{2}
            \bar{g}^{\alpha\beta}
            ( \bar{\nabla}_{\mu} h_{\nu\beta}
            + \bar{\nabla}_{\nu} h_{\mu\beta}
            - \bar{\nabla}_{\beta} h_{\mu\nu}).
\end{eqnarray}

We may consider only the variation of conformal factor
since the metric can be always written
in conformally
flat form where $h_{\mu\nu}(x^+,x^-)=2\delta\rho(x^+,x^-)
\bar{g}_{\mu\nu}(x^+,x^-)$ in two dimensions.
Also we keep ourselves in the Kruskal gauge, which tells us that
\begin{equation}
\label{eq:imp}
\delta\rho (x^+,x^-)= \delta\phi (x^+,x^-).
\end{equation}
Note that the classical background
solution (11) also satisfies the condition $\bar{\rho}=\bar{\phi}$
since the residual gauge symmetry can be fixed in the conformal gauge [3].
For the perturbed fields, the same condition holds, and the full
solution $(\bar{\phi} + \delta \phi)$
will be meaningful within the same gauge fixing condition.
In this gauge, the linearized equation (\ref{eq:lpeom})
can be written as
\begin{eqnarray}
\label{eq:lpeom++}
\partial_+^2 \delta\phi \hspace{-0.2cm}&-&\hspace{-0.2cm} 4 \partial_+
\bar{\phi} \partial_+ \delta \phi
=0, \\
\label{eq:lpeom--}
\partial_-^2 \delta\phi \hspace{-0.2cm}&-&\hspace{-0.2cm}4 \partial_-
\bar{\phi} \partial_- \delta \phi
=0,   \\
\label{eq:lpeomd}
\partial_+ \partial_- \hspace{-0.2cm} &\delta  \phi &\hspace{-0.2cm} - 2
\partial_+ \bar{\phi} \partial_- \delta \phi
- 2 \partial_- \bar{\phi} \partial_+ \delta \phi
- 2 (\partial_+ \partial_- \bar{\phi}
- 2 \partial_+ \bar{\phi} \partial_- \bar{\phi}) \delta \phi = 0,
\end{eqnarray}
where eqs.(\ref{eq:lpeom++}) and (\ref{eq:lpeom--}) are constraint equations,
and eq. (\ref{eq:lpeomd}) is a
perturbed dilaton equation.

To obtain perturbed solutions, we transform
equations of motion (17)-(19) from the Kruskal coordinates
$(x^+,x^-)$ to the tortoise coordinates $(t,r^\ast)$. In the $(t,r^\ast)$
coordinates, we can easily consider non-singular region outside the
black hole, and treat time $t$ explicitly.
The transformation is given by
\begin{eqnarray}
\label{eq:ct}
x^+ -x_0^+&=& +\frac{1}{\lambda} e^{+\lambda(t+r^\ast)} > 0, \nonumber \\
x^- -x_0^-&=& -\frac{1}{\lambda} e^{-\lambda(t-r^\ast)} < 0,
\end{eqnarray}
where $r^\ast$ ranges from $-\infty$ to $+\infty$, and the even horizon is
at $r^\ast \rightarrow -\infty$, and the infinity
$r^\ast \rightarrow +\infty$
corresponds to the asymptotically flat region.
We assumed that $\lambda$ is positive for simplicity.
Then, we rewrite the eqs. (17)-(\ref{eq:lpeomd})
in $(t,r^\ast)$ coordinates as
\begin{eqnarray}
\label{eq:lpeom++r}
(\partial_t + \partial_{r^\ast} )^2 \delta \phi
- 2 \lambda (\partial_t + \partial_{r^\ast} ) \delta \phi
+ \frac{4\lambda}{(1+\frac{m}{\lambda}e^{-2\lambda r^{\ast} })}
(\partial_t + \partial_{r^\ast} )  \delta \phi = 0,
\end{eqnarray}
\begin{eqnarray}
\label{eq:lpeom--r}
(\partial_t - \partial_{r^\ast} )^2 \delta \phi
+ 2 \lambda (\partial_t - \partial_{r^\ast} ) \delta \phi
+ \frac{4\lambda}{(1+\frac{m}{\lambda}e^{-2\lambda r^\ast })}
( - \partial_t + \partial_{r^\ast} )  \delta \phi = 0,
\end{eqnarray}
\begin{eqnarray}
\label{eq:lpeomdr}
(- \partial_t^2 + \partial_{r^{\ast}}^{2}) \delta \phi
+ \frac{4\lambda}{(1+\frac{m}{\lambda}e^{-2\lambda r^\ast})}
\partial_{r^\ast} \delta\phi
+ \frac{4\lambda^2}{(1+\frac{m}{\lambda}e^{-2\lambda r^\ast})}
\delta \phi = 0.
\end{eqnarray}
By adding and subtracting eqs. (\ref{eq:lpeom++r}) and (\ref{eq:lpeom--r}),
the following simplified relations are given,
\begin{eqnarray}
\label{eq:sub}
(\partial_t^2 +\partial_{r^\ast}^2) \delta \phi + 2 U(r^\ast) \partial_r^\ast
\delta \phi &=&0,  \\
\partial_t \partial_r^\ast \delta \phi + U(r^\ast) \partial_t \delta \phi &=&0,
\end{eqnarray}
where $U(r^\ast) \equiv \lambda\left( \frac{1-\frac{m}{\lambda}e^{-2\lambda
r^\ast}}
            {1+\frac{m}{\lambda}e^{-2\lambda r^\ast}} \right)$.

The constraint equation (25) is easily solved with two unknown functions as
\begin{eqnarray}
\label{eq:delphi}
\delta\phi(t \hspace{-0.5cm}&,&\hspace{-0.5cm} r^\ast)= A(r^\ast) + b(t)
e^{-\int^{t^\ast} dr^\ast U(r^\ast)}, \\
A(r^\ast)\hspace{-0.3cm}&=&\hspace{-0.3cm} e^{-\int^{r^\ast} dr^\ast U(r^\ast)}
                     \left( \int^{r^\ast} \{ a(r^\ast)
                            e^{\int^{r^\ast} dr^\ast U(r^\ast)}\} dr^\ast
\right), \nonumber \\
\int^{r^\ast}\hspace{-0.3cm} &dr^\ast& \hspace{-0.3cm} U(r^\ast) =\lambda
r^\ast
+\left( 1+\frac{m}{\lambda} e^{-2\lambda r^\ast}\right),  \nonumber
\end{eqnarray}
where $a(r^\ast)$ and $b(t)$ are integration functions in solving the
differential equation.
Interestingly the perturbed dilaton solution is composed of two functions;
one is dependent only on the space coordinate, and the other is
space-time dependent part which is reminiscent of some propagating modes.

For more details, plugging eq. (26) into the dilaton
equation (23) and eq. (\ref{eq:sub}), we obtain two separated
ordinary
differential equations,
\begin{eqnarray}
\frac{d^2}{d {r^\ast}^2} A(r^\ast) + 2 U(r^\ast) \frac{d}{d r^\ast} A(r^\ast)
                                                     &=&0,  \\
\frac{d^2}{d t^2}{b} (t) -\lambda^2 b(t) &=& 0.
\end{eqnarray}
These equations yield exact solutions as follows,
\begin{eqnarray}
A(r^\ast) &=& d -  \frac{c e^{-2\lambda r^\ast}}
               {(1+\frac{m}{\lambda}e^{-2\lambda r^\ast})}, \\
b(t) &=& \alpha^- ~e^{\lambda t}  + \alpha^+ ~ e^{-\lambda t},
\end{eqnarray}
where $c,~d,$ and $ \alpha^\pm $ are space-time independent constants.
{}From the boundary condition $A(\infty)=0$ for the asymptotically
flatness, and we set $ d=0$.
Note that in ref. [7,8], authors have qualitatively considered second part
$b(t)$ without amplitudes in our general solution (26) and neglected
the time-independent part $a(r^\ast)$ (or $A(r^\ast)$)
which arise from the integration with
respect to time in eq. (25).

Then, the linear perturbation of dilaton (graviton) field is given by
\begin{equation}
\delta \phi (t, r^\ast) = \frac{-c e^{-2\lambda r^\ast} +\alpha^- ~e^{\lambda
(t-r^\ast)}  + \alpha^+ ~e^{-\lambda (t+r^\ast)}}{(1+\frac{m}{\lambda}e^
{-2\lambda r^\ast}) }.
\end{equation}
At this stage, one might think that the black hole solution is unstable
against the time-dependent linear perturbation since the perturbed
dilaton solution has an exponentially growing mode with respect to time.
However, this is not the case because they corresponds to the variation of the
black hole position, which can be eliminated by taking an appropriate
coordinate as will be shown below.
In the Kruskal coordinates,
the solution (31) is written as
\begin{eqnarray}
\delta \phi (x^+, x^-) &=& \delta \rho (x^+ , x^-)   \nonumber \\
                       &=&   \frac{-(c +\lambda\alpha^- x_0^+ -\lambda \alpha^+
x_0^-)
                           + \lambda \alpha^- ~x^+
                         - \lambda \alpha^+ ~x^-}{ \left[\frac{m}{\lambda} -
                         \lambda^2 (x^+ - x_0^+)(x^- -x_0^-)\right]}.
\end{eqnarray}
Note that the linear perturbation is compatible with eq.
(\ref{eq:delphipm})
by comparing eqs. (32) and (12), we then identify
\begin{eqnarray}
c&=&\frac{\delta m}{2\lambda} +\frac{\lambda^2}{2}
               \left(\delta x_0^+ x_0^- + \delta x_0^- x_0^+ \right),        \\
\alpha^\pm &=& \pm \frac{\lambda}{2} \delta x_0^\pm   .
\end{eqnarray}
This identification shows that the physical origin of
unknown constants $\alpha^\pm$ in front
of the growing mode with time is due to the variation of position coordinates
of the black hole.

Furthermore, directly substituting the transformation rule (20) into
the black hole
solution (11), eq. (12) can be written as
\begin{equation}
\delta \phi (t, r^\ast)  = \frac{-\frac{\delta m}{2\lambda} e^{-2\lambda
r^\ast}}
{(1+\frac{m}{\lambda}e^
{-2\lambda r^\ast}) }
\end{equation}
only in terms of the mass parameter.
This is because the variation of black hole position is zero in the
tortoise coordinate.
Only when we allow
a constant shift as $x_0^\pm \rightarrow
x_0^\pm + a^\pm$ in eq. (20), then the growing modes with time
may exist in tortoise
coordinates. However, the black hole position (or variation of it)
are not gauge
invariant quantity and
just a location of black hole position which we could arbitrarily choose.
For this reason, $\alpha^\pm$ can be zero, and the growing modes
can be eliminated in the perturbed solution (31) and (32).

In summary, we have explicitly shown that the true local physical
degrees of freedom in the two-dimensioal dilaton gravity is zero
through the Hamiltonian constraint analysis.
Next, we have shown that in the two-dimensional dilaton gravity the only
possible physical perturbation (deformation) of the classical background
black hole is for the mass parameter, which is the global degree of freedom,
and the other two parameters are just a
position of black hole, which is not a gauge invariant physical
quantity. We have also considered from the direct linear perturbation around
the
static black hole, where we have obtained the explicit solution of
the linearized equations of motion and shown that the amplitudes in front of
the exponentially growing solution with time can be eliminated
by identifying them with the variation of black hole positions.
For these reasons, the two dimensional dilatonic black hole is stable.

\section*{Acknowledgements}
We would like to thank C. Lee for helpful discussions and encouragement.
W. T. Kim and J. Lee were supported in part by the Korea Science
and Engineering Foundation through the Center for Theoretical Physics (1994).
The present study was also supported in part by
the Basic Science Research Institute Program,
Ministry of Education, 1994, Project No. 2414.

\end{document}